# An S-matrix Formalism for the Nonclassical Optical Response of Plasmonic Sphere Aggregates

Xin Zheng, Christos Mystilidis, *Member*, *IEEE*, Christos Tserkezis, Guy A. E. Vandenbosch, *Fellow*, *IEEE*, Xuezhi Zheng, *Senior Member*, *IEEE*

*Abstract*—A computational method for the scattering of light by multiple nonclassical plasmonic nanospheres, each of which has multiple (non-)concentric dielectric or metallic layers, is presented. The electromagnetic (EM) response of the free electrons in the metals is described by three popular mesoscopic models: the nonlocal hydrodynamic Drude model (NLHDM) and its diffusive variant, namely the generalized nonlocal optical response (GNOR) model, as well as the surface response model (SRM). The main equation behind the method is set up by detailing the evaluation of the S-matrix for each individual spherical interface and the interactions amongst the interfaces. The algorithm is numerically validated by comparing with an in-house boundary element method (BEM) solver for a spherical NP with two smaller embedded spheres, and physically checked on a trimer configuration, where the responses from the NLHDM and SRM are contrasted with the local response model (LRM). In both cases a very good agreement is seen regarding frequency shifts and field enhancements.

Xin. Zheng is grateful for the China Scholarship Council (Grant No. 202206090027), China. Xuezhi Zheng, and Guy A. E. Vandenbosch are grateful for the C1 project (C14/19/083), the IDN project (IDN/20/014), the small infrastructure grant (KA/20/019) from KU Leuven and for G090017N and G088822N from the Research Foundation of Flanders (FWO). Xuezhi Zheng would like to also thank the support from the IEEE Antennas and Propagation Society Postdoctoral Fellowship, and for the FWO long travel grant, FWO V408823N. The Center for Polariton-driven Light – Matter Interactions (POLIMA) is funded by the Danish National Research Foundation (Project No. DNRF165) (Corresponding author: Xuezhi Zheng).

X. Zheng, C. Mystilidis, X. Zheng, and G. A. E. Vandenbosch are affiliated with the WaveCore Division, Department of Electrical Engineering (ESAT), KU Leuven, B-3001, Leuven, Belgium.
C. Tserkezis is affiliated with POLIMA – Center for Polariton-driven Light-Matter Interactions, University of Southern Denmark, 5230 Odense, Denmark.

Color versions of one or more of the figures in this article are available online at http://ieeexplore.ieee.org

## I. INTRODUCTION

IN classical electrodynamics, the electromagnetic (EM) response of metals is well described by a local macroscopic material parameter, e.g., $\chi$ (susceptibility), which can be a function of spatial position and frequency. The parameter links the polarization density with the electric field at *the same spatial point*, that is, $\mathbf{P}(\mathbf{r}) = \varepsilon_0 \chi(\mathbf{r}, \omega) \mathbf{E}(\mathbf{r})$ (where $\varepsilon_0$ is the vacuum permittivity and $\omega$ is the angular frequency; in this work an $\exp(-i\omega t)$ time dependence and the SI units are used) and marks the so-called local response model (LRM) [1], [2]. So far, the classical EM treatment of LRM has been applied to predict the optical response of many plasmonic nanostructures on the scale of tens of nm. However, the recent advances in nanofabrication technology have granted us access to the deep-nanometric regime (roughly from a fraction of 1 nanometer to a few nanometers [3]), where the confinement of EM waves has reached an unprecedented level [4], [5], and has enabled many ground-breaking applications, such as single-molecule level chemistry [6], [7], surface-enhanced Raman scattering [4], [5] enhanced light harvesting [8], [9], [10], and novel, beyond-CMOS optical circuits [11], to name a few. Together with this breakthrough, critical statements on the LRM have been made [2], [12], that is, the LRM, which puts all EM properties of metals in one macroscopic parameter, is not adequate to describe the physics at the deep-nm scale anymore. As remedies, many semiclassical models have been proposed [2]. The most popular ones are the nonlocal hydrodynamic Drude model (NLHDM) [13], its variant termed the generalized nonlocal optical response model (GNOR) [14], and the surface response model (SRM) [3], [15], [16]. In the NLHDM and GNOR, a gradient-divergence term added to the LRM can simulate the finite compressibility of the free electron gas and surface scattering phenomena, via the convective (NLHDM) and the convective-diffusive (GNOR) flow of free electrons in metals. Due to this term, the constitutive relation linking the electric field $\mathbf{E}(\mathbf{r})$ with the polarization current $\mathbf{P}(\mathbf{r})$ is not simply constructed at a single spatial point, but also concerns the neighborhood around it. This contrasts to the LRM, where only one spatial point is involved, i.e., the correlation is purely *local*. Hence, the NLHDM and GNOR are termed as *nonlocal response models.* On the other hand, instead of focusing on nonlocal corrections to the LRM at the equation of motion level, SRM lumps the nonclassical effects, e.g., the spill-out effect and surface scattering in the so-called Feibelman *d*-parameters for the metal–dielectric interface. The model splits the metal into two regions: a bulk region and a selvedge region, where the bulk still follows the LRM, while the behavior of fields crossing the selvedge region is described by a set of quantum-corrected boundary conditions (QC – BCs).

Accordingly, these emerging semiclassical models require a "rewriting" of established computational electromagnetics (CEM) algorithms. This nascent branch of CEM is termed computational mesoscopic electromagnetics (CMEM) [17], [18] and its task consists of numerically correctly predicting the nonclassical optical response from plasmonic nanostructures with deep-nm features, which are of both theoretical and experimental interest. So far, NLHDM, GNOR and SRM have been integrated with many CEM algorithms. For example, the differential equation-based methods, such as Finite Difference Time Domain (FDTD) [19], Finite Element Method (FEM) [20], discontinuous Galerkin method (DGTD) [21], [22] have been readily applied to study the nonclassical effects for arbitrary nanotopologies, for the NLHDM [23], [24], and for the SRM [3], [25] and, the integral equation-based methods, BEM [26], [27], discrete sources method (DSM) [28], [29] and Volumetric Method of Moments (V-MoM) [30], have been tailored to study NPs both in homogeneous space (for the NLHDM [26], [31], [32], [33] and for the SRM [34], [35]) and on a mirror (for the NLHDM [36]). Complementing such efforts, (semi-)analytical approaches, though limited to canonical (i.e., the planar, the cylindrical and the spherical) topologies, provide not only reference results for benchmarking the numerical algorithms, but also a proof-of-concept platform for revealing the physics behind. The interaction of light with individual canonical topologies was first solved in the context of physical discussions, e.g., NLHDM and GNOR (for cylinders and spheres) in [37], [38], [39] and the SRM (for spheres and planar layers) in [40], [41]. Further, these analytical treatments have evolved to more complicated cases, e.g., the transmission of light through multiple planar layers [42] and (concentric/non-concentric) cylindrical and spherical layers [43], [44], [45], [46], where one (or more) layers are nonclassical metals; and to cylinder and sphere on mirror [36], [47], where both the cylinder (sphere) and the mirror can be nonclassical.

In this work, we focus on the spherical case and develop an algorithm that treats the interaction of light with multiple nonoverlapping spherical interfaces (IFs) belonging to the same or multiple spheres (see **Fig. 1**) the inner region and/or outer region of which can be filled with nonclassical metals. This is in sharp contrast to previous efforts where solitary single or layered spheres are discussed [43], [45], [46], [48], [49]. In detail, we build the algorithm in two steps. As the first step, we derive the S-matrix for each spherical IF. The S-matrix is constructed by expanding the electric and magnetic fields in series of vector wave functions both inside and outside the IF and then matching the field quantities at the IF. Especially, for the NLHDM and GNOR, we distinguish three cases: (1) both the inner and the outer region of the IF are filled with local media; (2) the inner and the outer region are filled with local (nonlocal) medium and nonlocal (local) medium; (3) both the inner and the outer region are filled with nonlocal media. As the second step, the translation addition theorem is applied to describe the interactions between the IFs. Notably, for the NLHDM and GNOR, besides the translation of transverse waves in classical EM theory (e.g., see Appendix D in [50]), the translation of longitudinal waves may be also needed. Since the S-matrix is at the center of the developed algorithm, we name the algorithm as an "S-matrix" based algorithm. The text is organized as follows. In Section II, we briefly review the semiclassical material models and derive the S-matrices. Then, the main equation to be solved by the algorithm and the key theoretical element, namely the translation addition theorem, are presented. Then, to validate, in Section III, a spherical IF with two eccentric spherical cores is tested. The absorption cross sections and the near fields by the proposed algorithm are compared with the ones from an in-house developed BEM solver, where an excellent agreement is shown. Further, one physical check considering a sodium (Na) trimer is presented for LRM, SRM and NLHDM, where the results are in line with the previously reported physical findings.

## II. Theory

### A. The Nonlocal Hydrodynamic Drude Model (NLHDM) and the Surface Response Model (SRM)

First, we briefly discuss the NLHDM and its variant GNOR. Many detailed discussions on the two physical models can be found in our previous works [26], [27], [31], [32], [36], [43], [46], [47]. Both models introduce an additional partial differential equation (PDE) of the same form,

$$\xi^2 \nabla \left[ \nabla \cdot \mathbf{P}_f(\mathbf{r}) \right] + \mathbf{P}_f(\mathbf{r}) = -\varepsilon_0 \frac{\omega_p^2}{\omega(\omega + i\gamma)} \mathbf{E}(\mathbf{r}). \quad (1)$$

In (1), $\mathbf{P}_f$ is the free-electron polarization current density; $\omega_p$ and $\gamma$ are the plasma frequency and the damping rate of the free electron gas in a metal and, for the NLHDM, $\xi^2$ is,

$$\xi^2(\omega) = \frac{\beta^2}{\omega(\omega + i\gamma)}. \quad (2)$$

In (2), $\beta$ is a parameter related to the Fermi velocity and typically taken to its high-frequency limit $\sqrt{3/5}v_F$ [51]. Then, the first term of (1) reflects the quantum pressure which is generated by the Pauli exclusion principle. On the other hand, for the GNOR, $\xi$ reads,

$$\xi^2(\omega) = \frac{\beta^2}{\omega(\omega + i\gamma)} + \frac{D}{i\omega}. \quad (3)$$

The term containing $D$ in (3) is related to the diffusion effect that removes a high concentration of free electrons at the boundary of a nanoparticle. It is a phenomenological diffusion parameter and the recipes to obtain it can be found in [52]. If we neglect the $\xi$ term in (1), it reduces to the Drude model which involves only one spatial point $\mathbf{r}$. That is, the inclusion of the gradient–of–divergence in the $\xi$ term introduces nonlocality. The PDE in (1) enters Maxwell's equations via the electric displacement field,

$$\mathbf{D}(\mathbf{r}) = \varepsilon_0 \varepsilon_{bd} \mathbf{E}(\mathbf{r}) + \mathbf{P}_f(\mathbf{r}). \qquad (4)$$

$\varepsilon_{bd}$ is the (relative) bound electron permittivity of the medium (see [27], [53] for more detailed explanations). Due to the PDE in (1), additional boundary conditions (ABCs) next to the two conventional ones that assure the continuity of the tangential components of the electric and magnetic fields are needed. For an IF separating a medium described by LRM (that is, a *local* medium) and a medium described by NLHDM or GNOR (i.e., a *nonlocal* medium), the ABC reads,

$$\mathbf{n} \cdot \varepsilon_{1,bd} \mathbf{E}_1 = \mathbf{n} \cdot \varepsilon_2 \mathbf{E}_2. \qquad (5)$$

In (5), $\mathbf{n}$ is the unit normal of the IF, pointing from the inner region to the outer region; the subscripts 1 and 2 indicate the inner and outer regions separated by the IF; it is assumed that region 1 is filled with a *nonlocal* medium, $\varepsilon_{1,bd}$ means the bound electron permittivity of the medium in region 1, and region 2 is filled with a *local* medium with a permittivity $\varepsilon_2$. The ABC in (5), what is actually the so-called Sauter ABC (see [54]), suggests the termination of the free-electron polarization current at the IF, i.e., $\mathbf{n} \cdot \mathbf{P}_f = 0$.

However, when the IF separates a *nonlocal* medium from another *nonlocal* medium, two ABCs are needed,

$$\begin{aligned} \mathbf{n} \cdot \varepsilon_{1,bd} \mathbf{E}_1 &= \mathbf{n} \cdot \varepsilon_{2,bd} \mathbf{E}_2, \\ \frac{\beta_1^2}{\omega_{1,p}^2} \varepsilon_{1,bd} \nabla \cdot \mathbf{E}_1 &= \frac{\beta_2^2}{\omega_{2,p}^2} \varepsilon_{2,bd} \nabla \cdot \mathbf{E}_2. \end{aligned} \qquad (6)$$

Notably, the second equation of (6) emphasizes the continuity of charge across the IF (see detailed explanations on (6) in [54] and [55]).

Second, lying at the core of the SRM are two generalized, or quantum-corrected, boundary conditions (QC-BCs),

$$\mathbf{E}_2^\parallel - \mathbf{E}_1^\parallel = -d_\perp \cdot \nabla_\parallel \left( \mathbf{E}_2^\perp - \mathbf{E}_1^\perp \right). \qquad (7)$$

$$\mathbf{H}_2^\parallel - \mathbf{H}_1^\parallel = -i\omega d_\parallel \cdot \mathbf{n} \times \left( \mathbf{D}_2^\parallel - \mathbf{D}_1^\parallel \right). \qquad (8)$$

The SRM treats an IF between a metal and a dielectric. In this model, the two media are modeled by using the LRM. But, the boundary conditions are modified: as seen in (7) and (8), the BCs that ensure the continuity of $\mathbf{E}^\parallel$ and $\mathbf{H}^\parallel$, i.e., the tangential components of the electric and magnetic fields are broken. Instead, the terms on the right-hand side (RHS) of (7) and (8), which are related to the dipole moments of the induced charge distribution normal to and in-plane of the boundary of a metal, introduce discontinuities. $d_\parallel$ and $d_\perp$ are the so-called *Feibelman* parameters [40], [56] and can be extracted from LRM, NLHDM or even a full quantum mechanical calculation, e.g., Time-dependent Density Functional Theory [3], [57].

*B. Main Equation*

After discussing the material models, we next look at the main equation behind the proposed algorithm, and how the material models enter the main equation is discussed along the way. Consider the interaction of an incident field, which can be a plane wave, the radiation from a dipole or an electron beam, with $N$ multiple spherical IFs (see **Fig. 1**). The main equation that controls the multiple scattering process is,

$$\begin{pmatrix} \mathbf{p}_-^i \\ \mathbf{p}_+^o \end{pmatrix} = \mathbf{S}_p \cdot \left\{ \mathbf{p}^e + \sum_{q=1, q \neq p}^{N} \begin{pmatrix} \mathbf{T}_{pq}(i,+;i,-) & \mathbf{T}_{pq}(i,+;o,+) \\ \mathbf{T}_{pq}(o,-;i,-) & \mathbf{T}_{pq}(o,-;o,+) \end{pmatrix} \cdot \begin{pmatrix} \mathbf{q}_-^i \\ \mathbf{q}_+^o \end{pmatrix} \right\}. \qquad (9)$$

(9) is listed for the $p^{\text{th}}$ particle and the sum on the RHS of (9) covers all $N$ particles except for the $p^{\text{th}}$ particle. In (9), the key term is the S-matrix of the $p^{\text{th}}$ IF, i.e., $\mathbf{S}_p$, defined as,

$$\begin{pmatrix} \mathbf{p}_-^i \\ \mathbf{p}_+^o \end{pmatrix} = \mathbf{S}_p \cdot \begin{pmatrix} \mathbf{p}_+^i \\ \mathbf{p}_-^o \end{pmatrix}. \qquad (10)$$

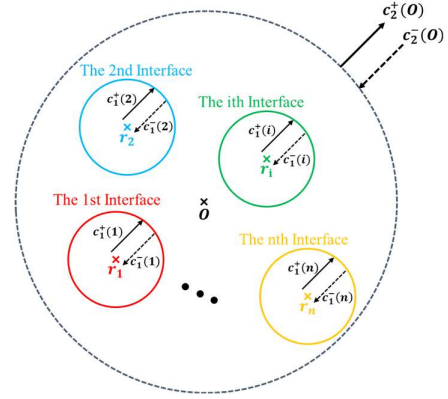

**Fig. 1.** An illustration of $N$ interacting nanospheres. In the figure, there are $N$ nanospheres. The subscripts 1 and 2 represent the inner and outer region of each IF, respectively. The arrows indicate whether the waves are converging or diverging with respect to the origin $O$. The superscripts "+" and "-" mark the wave amplitudes that are associated with the waves diverging along or converging against the radial direction. The dashed circle outside all nanospheres is an imaginary boundary collecting all the scattered and incident waves.

The S-matrix relates the expansion coefficients ($\mathbf{p}_+^i$ and $\mathbf{p}_-^o$) of the "incoming" field which is incident on the $p^{\text{th}}$ IF with the ones ($\mathbf{p}_-^i$ and $\mathbf{p}_+^o$) of the "outgoing" field which leaves the $p^{\text{th}}$ IF behind. Here, the superscripts $i$ and $o$ indicate inner and outer regions, i.e., the regions in which the field resides; and the subscripts "+" and "-" mark that the fields are traveling along or against the normal (pointing from the inner region to the outer region) of the IF. Here, we would like to emphasize that for either the inner or the outer region, the material can be *local* or *nonlocal*. For the *nonlocal* case, i.e., the most complicated case, the electric field can be expanded in spherical wave functions (see the detailed form of $\mathbf{M}_{nm}$, $\mathbf{N}_{nm}$ and $\mathbf{L}_{nm}$ in [50]) with respect to the center of the $p^{\text{th}}$ IF (i.e., $\mathbf{r}_p$ in (11)),

$$\begin{aligned} \mathbf{E}(\mathbf{r} - \mathbf{r}_p) &= \sum_{nm} \mathbf{M}_{nm}(k, \mathbf{r} - \mathbf{r}_p) \cdot a_{nm} \\ &+ \sum_{nm} \mathbf{N}_{nm}(k, \mathbf{r} - \mathbf{r}_p) \cdot b_{nm} \\ &+ \sum_{nm} \mathbf{L}_{nm}(\kappa, \mathbf{r} - \mathbf{r}_p) \cdot c_{nm}. \end{aligned} \qquad (11)$$

The summation in (11) is conducted with respect to $n$ and $m$ where $n$ is a *nonnegative* integer and $m$ is an integer between $-n$ and $+n$. Due to nonlocality [26], [27], a longitudinal wave $\mathbf{L}_{mn}$ with a longitudinal wavenumber $\kappa$ is included, in addition to the transverse waves $\mathbf{M}_{nm}$ and $\mathbf{N}_{nm}$ that are characterized by the transverse wavenumber $k$. For a *local* medium, *only* $\mathbf{M}_{nm}$ and $\mathbf{N}_{nm}$ exist and in this case $n$ is a positive integer. When considering a wave traveling along (against) the normal, the spherical Hankel

(Bessel) function should be used in the expansion. To be complete, **p** in (10) is a column vector composed of all the expansion coefficients $a_{nm}$, $b_{nm}$ and (if the medium is nonlocal) $c_{nm}$ of the spherical wave functions $\mathbf{M}_{nm}$, $\mathbf{N}_{nm}$ and $\mathbf{L}_{nm}$ (if the medium is nonlocal).

Further, four types of IFs can be formed due to the combinations of materials filling the inner and the outer region. First, the IF can separate a *local* medium (inner region) from a *local* medium (outer region). This applies to both the classical case and the SRM. For the latter, the QC-BCs in (7) and (8) must be applied to link the fields on both sides of the IF. Second, the IF can separate a *local* medium (inner region) from a *nonlocal* medium (outer region). Third, the IF can separate a *nonlocal* medium (inner) from a *local* medium (outer region). For the second and the third cases, the two conventional IF conditions and the ABC in (5) must be applied. Fourth, the IF can separate a *nonlocal* medium from another *nonlocal* medium. For this case, the two conventional IF conditions and the ABCs in (6) must be applied. It is noted that in the above when referring to a *nonlocal* medium, the EM response of the medium is described by the NLHDM, and in any case the NLHDM should *not* be used together with the SRM, in which the EM response of the bulk region is described by the LRM. To this end, the S matrix of the $p^{\text{th}}$ IF, i.e., $\mathbf{S}_p$ in (10), can be derived by matching the field expansions inside and outside of the IF by the relevant boundary conditions (see our previous works for the details on how the field matching leads to the S matrix [43]).

Next, different sources can contribute to the expansion coefficients of the "incoming" field (i.e., the RHS of (10)),

$$\begin{pmatrix} \mathbf{p}^i_+ \\ \mathbf{p}^o_- \end{pmatrix} = \mathbf{p}^e + \begin{pmatrix} \mathbf{p}^i_+ \\ \mathbf{p}^o_- \end{pmatrix}. \quad (12)$$

Regarding the 1st term on the RHS of (12), $\mathbf{p}^e$ represents the expansion coefficients of the incident field, which is imposed and is known in our calculation. The 2nd term accounts for the scattered field generated by all other $N-1$ IFs. On the one hand, the scattered field can be expanded with respect to the center of the $p^{\text{th}}$ spherical IF, i.e., $\mathbf{r}_p$. Accordingly, the expansion coefficients are $\mathbf{p}^i_+$ and $\mathbf{p}^o_-$. On the other hand, the scattered field can be expanded against the centers of the other $N-1$ IFs. For each IF, the coefficients are $\mathbf{q}^i_-$ and $\mathbf{q}^o_+$. Based on the addition theorem (see Appendix D in [50]), $\mathbf{p}^i_+$ and $\mathbf{p}^o_-$ can be linked with $\mathbf{q}^i_-$ and $\mathbf{q}^o_+$ as,

$$\begin{pmatrix} \mathbf{p}^i_+ \\ \mathbf{p}^o_- \end{pmatrix} = \sum_{q=1,q \neq p}^{N} \begin{pmatrix} \mathbf{T}_{pq}(i,+;i,-) & \mathbf{T}_{pq}(i,+;o,+) \\ \mathbf{T}_{pq}(o,-;i,-) & \mathbf{T}_{pq}(o,-;o,+) \end{pmatrix} \cdot \begin{pmatrix} \mathbf{q}^i_- \\ \mathbf{q}^o_+ \end{pmatrix}. \quad (13)$$

For the sake of conciseness, the detailed derivations for (13) are given in the Appendix. Here, we instead focus on the meaning of the $\mathbf{T}_{pq}$ matrices. First, the subscripts "$pq$" mark the fact that the "source" IF emitting the scattered field is the $q^{\text{th}}$ IF and the "observation" IF is the $p^{\text{th}}$ IF. Second, the 1st (3rd) variable in the brackets can take two values, either $i$ or $o$, indicating the inner region or the outer region of the $p^{\text{th}}$ ($q^{\text{th}}$) IF. The 2nd (4th) variable in the brackets can take two values, either "+" or "-", denoting the wave propagating along or against the unit radial vector of the $p^{\text{th}}$ ($q^{\text{th}}$) IF. Then, $\mathbf{T}_{pq}(i,+;i,-)$, as an example, describes how the wave traveling in the inner region of the $q^{\text{th}}$ IF and against the radial direction of the IF interacts the wave propagating in the inner region of the $p^{\text{th}}$ IF and along the radial direction of the relevant IF. Since the $p^{\text{th}}$ and the $q^{\text{th}}$ IF are not the same IF, they will never share a common inner region. Thus, by our discussions in Appendix, $\mathbf{T}_{pq}(i,+;i,-)$ is always a zero matrix. The dimensions of the matrix are determined by the sizes of $\mathbf{p}^i_+$ and $\mathbf{q}^i_-$ where it is noted that the size of $\mathbf{p}^i_+$ ($\mathbf{q}^i_-$) depends on whether the inner region of the $p^{\text{th}}$ ($q^{\text{th}}$) IF is filled by a *local* or a *nonlocal* medium and the NLHDM is employed. For the latter case, the extra expansion coefficients $c_{nm}$ must be accounted for. The above discussions can be readily generalized to explain $\mathbf{T}_{pq}(i,+;o,+)$, $\mathbf{T}_{pq}(o,-;i,-)$ and $\mathbf{T}_{pq}(i,+;i,-)$ (see **Fig. A1** and its caption for more information).

At last, we combine (10) - (13) and arrive at the main equation in (9). As such, we can list (9) for all $N$ IFs. After solving the system (see below for the details of our coding in MATLAB), the expansion coefficients of the "outgoing" field for all the particles, i.e., $\mathbf{p}^i_-$ and $\mathbf{p}^o_+$ where $p$ ranges from 1 to $N$, can be obtained. By using these coefficients in the field expansion in (11) and together with the known incident field, we can, in principle, reconstruct the field everywhere in space.

### III. RESULTS

In this section, we demonstrate numerical results from the proposed algorithm. Here, we raise two examples. The first is about a core-shell structure (see **Fig. 2**) where two spherical (eccentric) cores are included in a spherical IF. The second concerns a trimer structure (see **Fig. 5**). For the former, we compare the results from the proposed algorithm with data generated by an in-house developed tool, while the latter emphasizes discussions from a more physical perspective. The algorithm is coded in MATLAB, and the computation is performed on a workstation with AMD 7950X (16 cores – 32 threads CPU) and 128 GB RAM. For the sake of completeness, when solving the system formed by (9), we use "equilibrate" in MATLAB to permute and rescale the system matrix, so that the efficiency and stability of the GMRES iterative solver is improved.

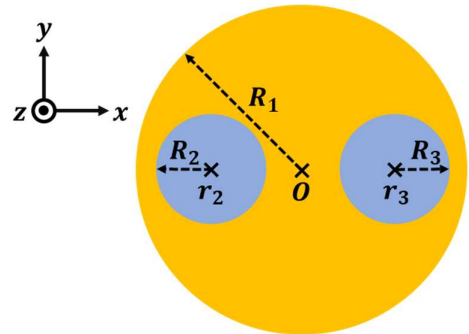

**Fig. 2.** Illustration of a core-shell structure in vacuum. The structure has three spherical IFs. The outermost one is centered at $O$ (the origin of the coordinate system) and its radius $R_1$ is 10 nm. The two inner IFs are centered at $\mathbf{r}_2 = (-5,0,0)$ nm and $\mathbf{r}_3 = (+5,0,0)$ nm.

## A. Core-Shell Structure

First, we consider a core-shell structure as an example (see **Fig. 2**). The structure is placed in vacuum and has three spherical IFs. The outer IF separates gold (Au, the inner region) from vacuum (the outer region). The center of the outer IF is set as the origin $O$ of the coordinate system and its radius is $R_1 = 10$ nm. The inner IFs separate silver (Ag, the inner region) from Au (the outer region). The centers of the two IFs are located at $\mathbf{r}_2 = (+5,0,0)$ nm and $\mathbf{r}_3 = (-5,0,0)$ nm, respectively, and their radii $R_2$ and $R_3$ are 4 nm. For the NLHDM, the plasma frequency, the damping rate and the bound electron permittivity of both Au and Ag are taken from tabulated data [58], and the Fermi velocity of Au and Ag is $1.40 \times 10^6$ m/s and $1.39 \times 10^6$ m/s [59]. The structure is excited by a plane wave travelling along the positive $z$ direction and polarized along the $x$ axis. The amplitude of the incident electric field is 1 V/m. We consider a wavelength range from 400 nm to 700 nm with 31 sampling points taken in between. As the result of the simulations, the absorption cross sections $\sigma_{abs}$ are evaluated. For this, the fields are collected on a sphere closely inside the outermost IF (the radius of the observation sphere is 9.9 nm) and the sphere is discretized by 1271 triangles. The fields and thus the Poynting vectors are evaluated at the centroids of all the triangles, and the absorbed power and the cross sections are obtained by numerical integrations. Our results are then compared with simulations using in-house BEM solver [26].

We consider three hypothetical cases: 1) both Ag and Au are treated as local media; 2) Ag is treated as a local medium, while Au is treated as a nonlocal medium; 3) both Ag and Au are treated as nonlocal media. Please note that case 2 is purely hypothetical and is raised only for the purpose of validation. That is, in a real-life scenario, the material model describing the medium inside and outside of an IF must be consistent (either both *local* or both *nonlocal*). The obtained absorption cross sections are shown in **Fig. 3**. It can be immediately observed from **Fig. 3**(a) – (c) that the simulated results demonstrate good agreements. A further evaluation of relative errors,

$$\text{err} = \frac{\sigma_{abs}^S - \sigma_{abs}^{BEM}}{\sigma_{abs}^S}, \quad (14)$$

leads to a maximum error within 4% (see **Fig. 3**(d) – (f)). It is seen that a larger relative error appears at short wavelengths, which is due to the mesh used in BEM. We have conducted BEM calculations with denser meshes (not shown here) and observe a better agreement.

In addition, we plot the amplitude of the near field at the resonances (see **Fig. 4**). For the three different cases, the main resonances are at 480 nm, 470 nm and 470nm, respectively. It is noted that a blue shift is observed for the *nonlocal* case which agrees with the previous results from physical discussions [47]. Then, a cut is taken on the $x - y$ plane and the circular cut has a radius of 10 nm. The cut is then triangulated, where the maximum area of the triangles is 0.311 nm$^2$ and the maximum radius of their inscribed circles is 0.244 nm, and the electric fields are evaluated at the centroids of all triangles. It can be seen from **Fig. 4** that the results from both the proposed method and the BEM agree well; for the near field, for the local-local case, the maximum relative error is 18.998%, and the average relative error is 0.828%; for the local-nonlocal case, the maximum relative error is -34.59%, and the average relative error is 0.915%; for the nonlocal-nonlocal case, the maximum relative error is 19.825%, and the average relative error is 0.762%. The maximum relative errors are observed at spatial points with relatively small field values. Also, the characteristic nonclassical effects can be observed, such as the reduced field enhancement in the gap between two inner IFs. For the local-nonlocal case, the appearing shells around the Ag spheres are due to the fact that Au is set as a nonlocal medium. There, the almost null fields are very likely due to the ABC in (5) which forces a vanishing normal component of the free electron polarization current.

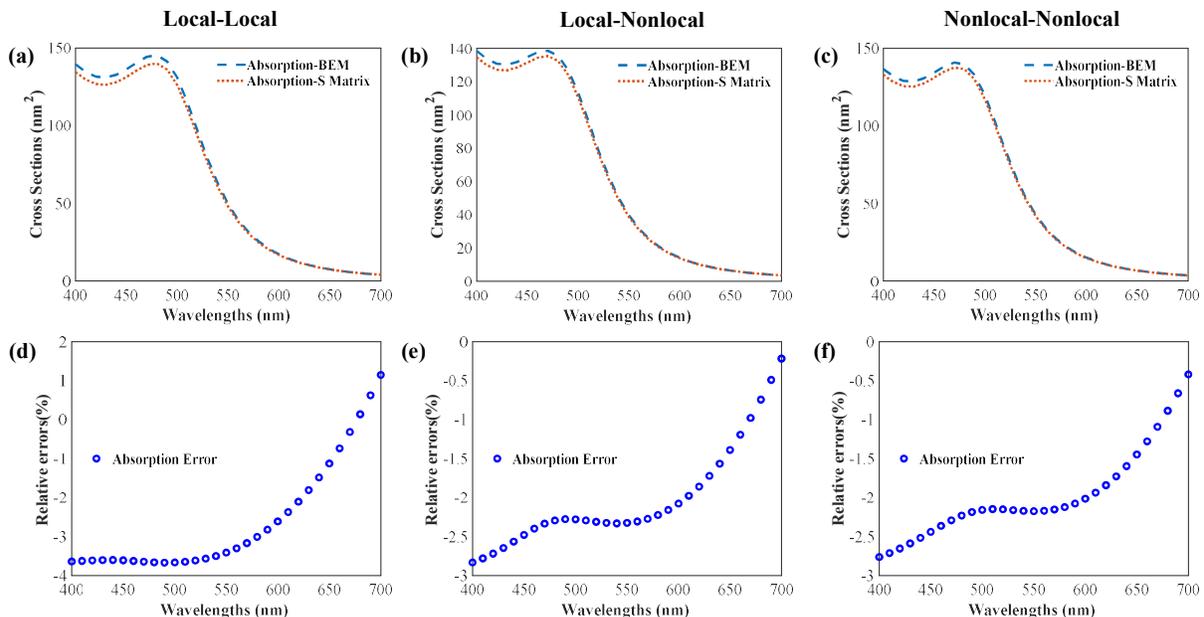

**Fig.3.** Evaluated absorption cross sections. (a) – (c) correspond to the three cases: the *local – local* case, the *local – nonlocal* case and the *nonlocal – nonlocal* case. In (a) – (c), the blue dashed lines represent the results from BEM, while the orange dotted lines represent the results from the proposed algorithm. In (d) – (f), the relative errors as defined in (14) are shown.

For the sake of completeness, the runtimes for the most complicated *nonlocal - nonlocal* case are compared. It takes BEM 1369.004 seconds to finish, while the runtime for the proposed method is 477.301 seconds. As a comment, in our calculation, the proposed method uses $n$ up to 19, while the BEM uses 878 and 508 triangles to discretize the outermost IF and the two inner IFs, respectively. For both cases, the number of spherical wave functions and the number of triangles is high enough, so that the proposed method and the BEM solver individually converge.

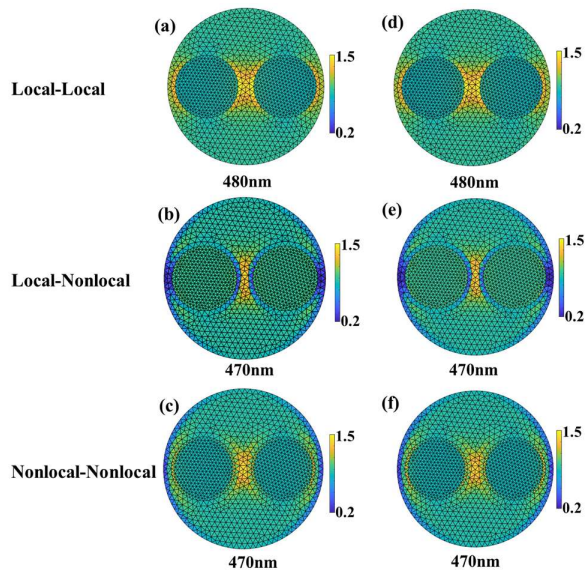

**Fig. 4.** Illustration of the electric field distribution along a cut made on the *x-y* plane. (a) – (c) are obtained from the proposed method, while (d) – (f) are from the in-house BEM. The three rows correspond to the three cases studied in **Fig. 3** In the figure, the colors are coded from blue to yellow to denote the magnitude of the fields. The same color code will be used for the rest of the text.

*B. Trimer*

Further, we compare the optical responses of the LRM, the NLHDM, and the SRM for a trimer structure (see **Fig. 5**). The structure is placed in vacuum and has three spherical IFs. Each IF separates sodium (Na, the inner region) from vacuum (the outer region). The origin $O$ of the coordinate system is set at the middle of two IFs parallel to the $x$ axis (see **Fig. 5**), and the radius of each sphere is $R = 10$ nm. To clearly show the effects of the LRM, the NLHDM and the SRM, five gap sizes are selected between IFs and IF gaps $d$'s change from 1 nm to 5 nm with 1 nm as a step. For the NLHDM, the plasma frequency, the damping rate and the bound electron permittivity of Na are taken from data in [48], and the Fermi velocity of Na is $1.06 \times 10^6$ m/s. The structure is excited by a plane wave travelling along the positive $z$ direction and polarized along the $x$ axis. The amplitude of the incident electric field is 1 V/m. Here, we consider a wavelength range from 200 nm to 700 nm. Here, 251 wavelength points are sampled to accurately locate the main resonance wavelengths (shown in **Fig. 6**).

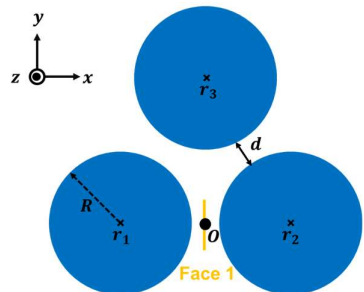

**Fig. 5.** Illustration of a trimer structure in vacuum. The structure has three spherical IFs. The origin $O$ of the coordinate system is set at the middle of the IF centered at $\mathbf{r}_1$ and the IF centered at $\mathbf{r}_2$, and the radius of each sphere is $R = 10$ nm. The gap $d$ between each nanoparticle changes from 1 nm to 5 nm with 1 nm as a step. Face 1 marks a cut in the $y - z$ plane, whose size is 10 nm by 10 nm.

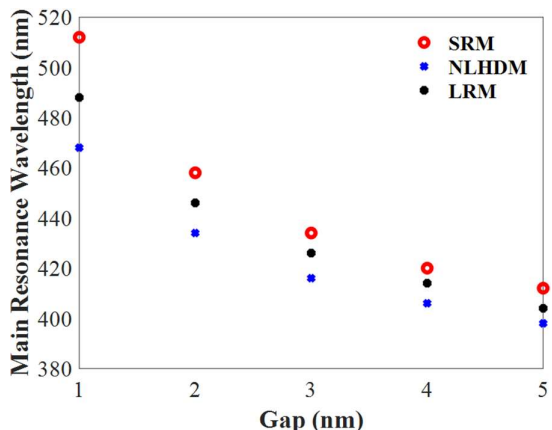

**Fig. 6.** A contrast of the wavelengths of the main resonance peaks in the absorption spectra for the LRM (black stars), the NLHDM (blue crosses) and the SRM (red circles) when the gap between the IFs changes from 1 nm to 5 nm. For the SRM, the main resonant wavelengths are 512 nm, 458 nm, 434 nm, 420 nm and 412 nm; for the NLHDM, the main resonant wavelengths are 468 nm, 434 nm, 416 nm, 406 nm and 398 nm; and for the LRM, the main resonant wavelengths are 488 nm, 446 nm, 426 nm, 414 nm and 404 nm.

In our calculations, the absorption cross sections for the five gap sizes are first obtained by integrating the Poynting vectors on three spheres inside the targeted IFs (the radius of each observation sphere is 9.9 nm). The wavelengths of the main resonances for the five cases are extracted. As a remark, it should be noted that, for the SRM, the most rigorous way to obtain the absorption cross sections is to set up the observation sphere just outside the three spheres. This is due to the jump in the tangential components of the fields introduced by the $d$ parameters in the QC – BC (see (7) and (8)). However, in our calculations, our main objective is to find the spectral position of the main resonance, which is not much affected by whether the observation sphere is placed just inside or outside.

When the gap is 1 nm and at the obtained main resonance wavelength, the corresponding near field distribution of the trimer structure is collected on a cut along the $xy$ plane at $z = 0$ with a size of 50 nm by 50 nm and the cut is discretized by 11977 triangles, where the maximum area of the triangles is 0.292 nm$^2$ and the maximum radius of their inscribed circles is 0.237 nm. Meanwhile, the electric field distributions are also collected on the cut in the $yz$ plane with a size of 10 nm by 10 nm, (see the yellow line in **Fig. 5**) located in the mid between the IF centered at $\mathbf{r}_1$ and the IF centered at $\mathbf{r}_2$, and discretized by 2737 triangles, where the maximum area of the triangles is 0.048 nm$^2$ and the maximum radius of their inscribed circles is 0.096 nm.

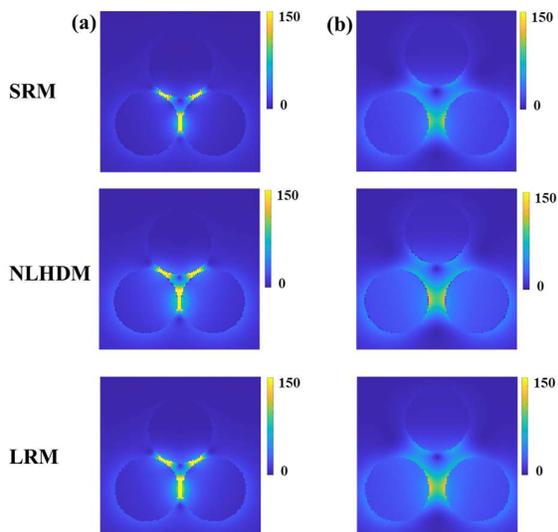

**Fig. 7.** Near-field mapping of the trimer structure for the SRM, the NLHDM and the LRM, respectively, at the main resonant frequencies for the case that the gap size is 1 nm (see (a)) and for the case that the gap size is 5 nm (see (b)). The magnitude of the electric field is plotted on a cut (the cut is on the $xy$ plane) with the size of 50 nm by 50 nm.

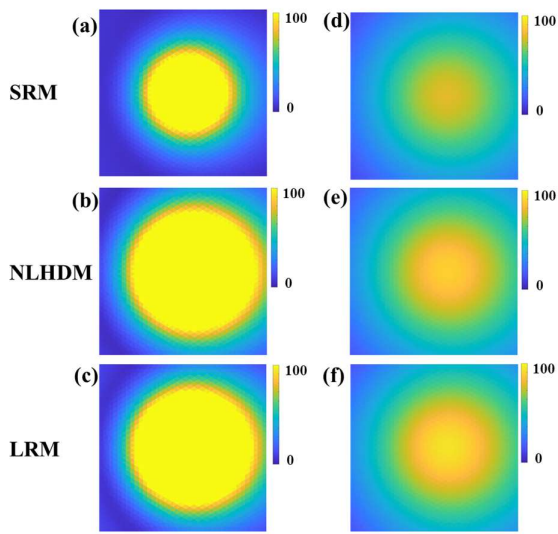

**Fig. 8.** A near-field mapping at Face 1 (see **Fig. 5**) for the SRM (the first row), for the NLHDM (the second row) and the LRM (the third row). The magnitude of the electric field distribution is evaluated for the case where the gap size is 1 nm (see (a), (b) and (c)) and for the case where the gap size is 5 nm (see (d), (e) and (f)).

The simulations are conducted in the proposed algorithm. To validate, we consider three cases: 1) Na is treated as a local medium; 2) Na is treated as a nonlocal medium whose dynamics is controlled by the NLHDM; 3) Na is treated as a local medium, supplemented by the SRM.

It is seen from **Fig. 6** that the main resonant wavelengths predicted by the NLHDM are blue-shifted from the ones by the LRM, while the optical response predicted by the SRM is redshifted from the one obtained by the LRM. For the gap size equal to 1 nm, for the former, the shift is 20 nm; while, for the latter, the shift is 24 nm. The spectral shift reduces when the gap increases: when the gap increases up to 5 nm, the spectral difference is smaller than 8 nm. If we continue increasing the gap size, the results obtained from the NLHDM and the SRM converge to those by LRM. The observed blueshift results from the "spill-in" nature of the NLHDM which effectively reduces the electrical size of the IF, leading to an increased gap size. On the contrary, the SRM considers a small transition region between Na and vacuum and accounts for the "spill-out" effects and thus increases the electrical size of the sphere, causing the redshift in the spectrum. Lastly, we consider the near field of the trimer structure for the LRM, the NLHDM and the SRM (see **Fig. 7** and **Fig. 8**) at the main resonances for the cases where the gap size is 1 nm and the gap size is 5 nm. First, the magnitude of the electric field is collected on the $xy$ plane. Comparing the 1 nm case with the 5 nm case, the maximum amplitude of the field decreases from 521.35 V/m to 136.42 V/m, for the LRM, from 424.03 V/m to 127.56 V/m for the NLHDM, and from 459.69 V/m to 115.79 V/m for the SRM. The reduced field amplitude is due to the weakened coupling of localized surface plasmons as the gap size increases. Apart from that, both the NLHDM and the SRM remove the surface charge concentration on the IF's boundary. The SRM further accounts for additional damping mechanisms, diminishing the field enhancement. Therefore, the field enhancement in the gap region, for the 1 nm case, reduces from 521.35 (for the LRM) to 424.03 (for the NLHDM) and 459.69 (for the SRM). These observations can be readily also confirmed by the field map evaluated on Face 1 (as marked in **Fig. 5**).

## IV. Conclusion

In conclusion, this work proposes a powerful algorithm based on the concept of S-matrix. The algorithm fully accounts for (1) the NLHDM and its GNOR variant, as well as the SRM, which are popular semiclassical models correcting the classical LRM; (2) the accompanying four material configurations inside and outside an IF: the *local – local* case, the *nonlocal – local* case, the *local – nonlocal* case and the *nonlocal – nonlocal* case (for the NLHDM); (3) nonconcentric spherical IFs. The proposed method is then validated against an in-house developed BEM solver. Good agreements are seen in both the absorption spectra and the near field mapping. A physical check is performed on a trimer sodium structure for the LRM, the NLHDM and the SRM, the results of which agree well with previous physical findings. The proposed method serves as a generalized platform for observing nonclassical optical response from multiple plasmonic spheres, adds to the list of works in CMEM, and can be a powerful tool in the study of mesoscopic electrodynamics.

APPENDIX: INTERACTION BETWEEN NON-CONCENTRIC INTERFACES

First, we look at two interacting spherical interfaces (see **Fig. A1**), e.g., the $p^{\text{th}}$ IF and the $q^{\text{th}}$ IF. As in (11), for the $p^{\text{th}}$ IF, the expansion is done with respect to $\mathbf{r}_p$, while for the $q^{\text{th}}$ IF it is done with respect to $\mathbf{r}_q$. To reconcile different expansion centers, we need to invoke the translation addition theorem for spherical waves (see Appendix D of [50]). For the case when two IFs are put in a region filled by a *local* medium, we have,

$$\mathbf{M}_{nm}(k, \mathbf{r}-\mathbf{r}_p) = \sum_{n'm'} \mathbf{M}_{n'm'}(k, \mathbf{r}-\mathbf{r}_q) A_{n'm',nm}(k, \mathbf{r}_{pq})$$
$$+ \sum_{n'm'} \mathbf{N}_{n'm'}(k, \mathbf{r}-\mathbf{r}_q) B_{n'm',nm}(k, \mathbf{r}_{pq}),$$

(15)

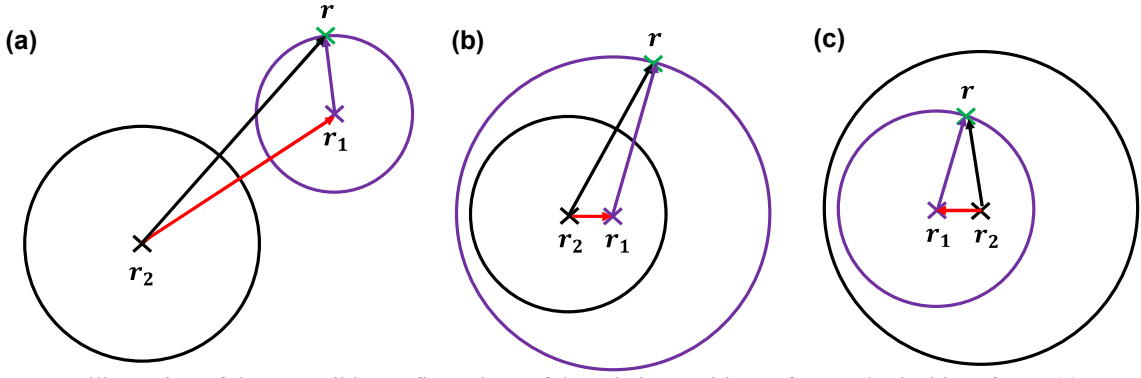

**Fig. A1.** An illustration of three possible configurations of the relative positions of two spherical interfaces. (a) Two IFs are in the outer region of each other; (b) The first IF (centered at $\mathbf{r}_1$) is in the outer region of the second IF (centered at $\mathbf{r}_2$); and (c) The first IF (centered at $\mathbf{r}_1$) is in the inner region of the second IF (centered at $\mathbf{r}_2$). The first IF is always marked by the purple color, while the second IF is done by the black color. The first IF and the second IF are the $p^{\text{th}}$ IF and the $q^{\text{th}}$ IF in the general discussions.

$$\mathbf{N}_{nm}(k, \mathbf{r}-\mathbf{r}_p) = \sum_{n'm'} \mathbf{M}_{n'm'}(k, \mathbf{r}-\mathbf{r}_q) B_{n'm',nm}(k, \mathbf{r}_{pq})$$
$$+ \sum_{n'm'} \mathbf{N}_{n'm'}(k, \mathbf{r}-\mathbf{r}_q) A_{n'm',nm}(k, \mathbf{r}_{pq}).$$

(16)

If the two IFs are in a region filled with a *nonlocal* medium, the longitudinal wave needs to be accounted for as well,

$$\mathbf{L}_{nm}(\kappa, \mathbf{r}-\mathbf{r}_p) = \sum_{n'm'} \mathbf{L}_{n'm'}(\kappa, \mathbf{r}-\mathbf{r}_q) \cdot \alpha_{n'm',nm}(\kappa, \mathbf{r}_{pq}).$$

(17)

In (15) - (17), $\mathbf{r}_{pq}$ is short for $\mathbf{r}_p - \mathbf{r}_q$. The mathematical forms of $A_{n'm',nm}$, $B_{n'm',nm}$ and $\alpha_{n'm',nm}$ can be found in Appendix D of [50] and will not be repeated here. It is noted that the transverse and longitudinal wavenumbers, i.e., $k$ and $\kappa$, are included in $A_{n'm',nm}$, $B_{n'm',nm}$ and $\alpha_{n'm',nm}$ via $z_{n''}$, where $z$ is either the spherical Bessel function (SBF) or the spherical Hankel function (SHF) and $n''$ is determined by $n$ and $n'$ together (see the rules in Appendix D of [50]). In detail, when $|\mathbf{r}_{pq}| > |\mathbf{r} - \mathbf{r}_q|$ (see **Fig. A1**(a)), the SBF is used in $\mathbf{M}_{nm}$, $\mathbf{N}_{nm}$ and $\mathbf{L}_{nm}$ and the SHF is used in $\mathbf{M}_{n'm'}$, $\mathbf{N}_{n'm'}$, $\mathbf{L}_{n'm'}$ and in $A_{n'm',nm}$, $B_{n'm',nm}$ and $\alpha_{n'm',nm}$. When $|\mathbf{r}_{pq}| < |\mathbf{r} - \mathbf{r}_q|$, the SBF is used in $A_{n'm',nm}$, $B_{n'm',nm}$ and $\alpha_{n'm',nm}$; but, in $\mathbf{M}_{nm}$, $\mathbf{N}_{nm}$ and $\mathbf{L}_{nm}$ and $\mathbf{M}_{n'm'}$, $\mathbf{N}_{n'm'}$ and $\mathbf{L}_{n'm'}$, either the SBF (see **Fig. A1**(b)) or the SHF (see **Fig. A1**(c)) is used.

Second, we take the two interacting spherical IFs in **Fig. A1**(a) as an example to show the construction of the relation in (13). This example can be generalized to other cases (e.g., **Fig. A1**(b) and (c) and $N$ interacting IFs). Here, we especially focus on the $1^{\text{st}}$ IF (e.g., see **Fig. A1**(a) in purple) and see how the scattered field due to the $2^{\text{nd}}$ IF (e.g., see **Fig. A1**(a) in black) can contribute to the incident field at the $1^{\text{st}}$ IF. Here, we assume the most complicated case, i.e., the $1^{\text{st}}$ IF is enveloped by a *nonlocal* medium. On the one hand, the scattered field $\mathbf{E}_s$ due to the $2^{\text{nd}}$ IF can be expanded with respect to the center of the $1^{\text{st}}$ IF, i.e., $\mathbf{r}_1$,

$$\mathbf{E}_s(\mathbf{r}-\mathbf{r}_1) = \sum_{n'm'} \mathbf{M}^{(1)}_{n'm'}(k, \mathbf{r}-\mathbf{r}_1) \cdot a_{n'm'}(1)$$
$$+ \sum_{n'm'} \mathbf{N}^{(1)}_{n'm'}(k, \mathbf{r}-\mathbf{r}_1) \cdot b_{n'm'}(1) \quad (18)$$
$$+ \sum_{n'm'} \mathbf{L}^{(1)}_{n'm'}(\kappa, \mathbf{r}-\mathbf{r}_1) \cdot c_{n'm'}(1),$$

In (18), $\mathbf{r}$ is a point on the $1^{\text{st}}$ IF; the superscript "1" in the wave functions is to mark that the SBF is used, since, for the $1^{\text{st}}$ IF, $\mathbf{E}_s$ due to the $2^{\text{nd}}$ is seen as a wave traveling towards the $1^{\text{st}}$ IF; and $a_{n'm'}(1)$, $b_{n'm'}(1)$ and $c_{n'm'}(1)$ denote that the expansion coefficients are related to the $1^{\text{st}}$ IF. On the other hand, the scattered field at the same point $\mathbf{r}$ due to the $2^{\text{nd}}$ IF can be expanded with respect to the center of the $2^{\text{nd}}$ IF, i.e., $\mathbf{r}_2$,

$$\mathbf{E}_s(\mathbf{r}-\mathbf{r}_2) = \sum_{nm} \mathbf{M}^{(3)}_{nm}(k, \mathbf{r}-\mathbf{r}_2) \cdot a_{nm}(2)$$
$$+ \sum_{nm} \mathbf{N}^{(3)}_{nm}(k, \mathbf{r}-\mathbf{r}_2) \cdot b_{nm}(2) \quad (19)$$
$$+ \sum_{nm} \mathbf{L}^{(3)}_{nm}(\kappa, \mathbf{r}-\mathbf{r}_2) \cdot c_{nm}(2).$$

In (19), the superscript "3" marks that the SHF is used in the spherical wave functions. By the addition theorem, $\mathbf{M}^{(3)}$, $\mathbf{N}^{(3)}$ and $\mathbf{L}^{(3)}$ centered at $\mathbf{r}_2$ can be expanded in terms of $\mathbf{M}^{(1)}$, $\mathbf{N}^{(1)}$ and $\mathbf{L}^{(1)}$ centered at $\mathbf{r}_1$,

$$\mathbf{M}_{nm}^{(3)}(k,\mathbf{r}-\mathbf{r}_2) = \sum_{n'm'} \mathbf{M}_{n'm'}^{(1)}(k,\mathbf{r}-\mathbf{r}_1) A_{n'm',nm}^{31}(k,\mathbf{r}_{21})$$
$$+ \sum_{n'm'} \mathbf{N}_{n'm'}^{(1)}(k,\mathbf{r}-\mathbf{r}_1) B_{n'm',nm}^{31}(k,\mathbf{r}_{21}), \quad (20)$$

$$\mathbf{N}_{nm}^{(3)}(k,\mathbf{r}-\mathbf{r}_2) = \sum_{n'm'} \mathbf{M}_{n'm'}^{(1)}(k,\mathbf{r}-\mathbf{r}_1) B_{n'm',nm}^{31}(k,\mathbf{r}_{21})$$
$$+ \sum_{n'm'} \mathbf{N}_{n'm'}^{(1)}(k,\mathbf{r}-\mathbf{r}_1) A_{n'm',nm}^{31}(k,\mathbf{r}_{21}). \quad (21)$$

$$\mathbf{L}_{nm}^{(3)}(\kappa,\mathbf{r}-\mathbf{r}_2) = \sum_{n'm'} \mathbf{L}_{n'm'}^{(1)}(\kappa,\mathbf{r}-\mathbf{r}_1) \alpha_{n'm',nm}^{31}(\kappa,\mathbf{r}_{21}). \quad (22)$$

We use the expansions in (20) - (22) in (19),

$$\mathbf{E}_s(\mathbf{r}-\mathbf{r}_2) = \sum_{nm}\sum_{n'm'} \mathbf{M}_{n'm'}^{(1)}(k,\mathbf{r}-\mathbf{r}_1) \cdot A_{n'm',nm}^{31}(k,\mathbf{r}_{21}) a_{nm}(2)$$
$$+ \sum_{nm}\sum_{n'm'} \mathbf{M}_{n'm'}^{(1)}(k,\mathbf{r}-\mathbf{r}_1) \cdot B_{n'm',nm}^{31}(k,\mathbf{r}_{21}) b_{nm}(2)$$
$$+ \sum_{nm}\sum_{n'm'} \mathbf{N}_{n'm'}^{(1)}(k,\mathbf{r}-\mathbf{r}_1) \cdot B_{n'm',nm}^{31}(k,\mathbf{r}_{21}) a_{nm}(2)$$
$$+ \sum_{nm}\sum_{n'm'} \mathbf{N}_{n'm'}^{(1)}(k,\mathbf{r}-\mathbf{r}_1) \cdot A_{n'm',nm}^{31}(k,\mathbf{r}_{21}) b_{nm}(2)$$
$$+ \sum_{nm}\sum_{n'm'} \mathbf{L}_{n'm'}^{(1)}(\kappa,\mathbf{r}-\mathbf{r}_1) \cdot \alpha_{n'm',nm}^{31}(\kappa,\mathbf{r}_{21}) c_{nm}(2). \quad (23)$$

Since the scattered field at the point $\mathbf{r}$ is unique, a comparison of (23) with (18) leads to,

$$a_{n'm'}(1) = \sum_{nm} A_{n'm',nm}^{31}(k,\mathbf{r}_{21}) \cdot a_{nm}(2)$$
$$+ \sum_{nm} B_{n'm',nm}^{31}(k,\mathbf{r}_{21}) \cdot b_{nm}(2), \quad (24)$$

$$b_{n'm'}(1) = \sum_{nm} B_{n'm',nm}^{31}(k,\mathbf{r}_{21}) a_{nm}(2)$$
$$+ \sum_{nm} A_{n'm',nm}^{31}(k,\mathbf{r}_{21}) b_{nm}(2), \quad (25)$$

$$c_{n'm'}(1) = \sum_{nm} \alpha_{n'm',nm}^{31}(\kappa,\mathbf{r}_{21}) c_{nm}(2). \quad (26)$$

In (20) - (26), the superscript "31" over $A$, $B$ and $\alpha$ emphasizes that the relations link the radiating-type spherical wave functions (SWF) with the standing-wave type SWF. Further, due to the relative position of the $2^{nd}$ and the $1^{st}$ IFs which implies $|\mathbf{r}_{21}| > |\mathbf{r}-\mathbf{r}_1|$, the SHF is used in $A$, $B$ and $\alpha$. We can collect the coefficients $a_{nm}(1)$, $b_{nm}(1)$ and $c_{nm}(1)$ in a column vector. As discussed in the main text, this column vector should be named as $\mathbf{1}_-^o$. Once again, $o$ underlines that the expansion coefficients correspond to the waves in the outer region of the $1^{st}$ IF, while $-$ denotes that the wave propagates against the radial direction of the IF. Also, in a similar way, we can put $a_{nm}(2)$, $b_{nm}(2)$ and $c_{nm}(2)$ in a column vector $\mathbf{2}_+^o$. Here, $o$ tells us that the coefficients are for the expansion of the waves in the outer region of the $2^{nd}$ IF and $+$ that the wave is traveling away from the IF. Then, we can rewrite (26) in a matrix form, i.e.,

$$\mathbf{1}_-^o = \mathbf{T}_{12}(o,-;o,+) \cdot \mathbf{2}_+^o. \quad (27)$$

In (27), we have used the symbol $\mathbf{T}_{12}(o,-;o,+)$ introduced in the main text. Again, the subscript "12" says that the matrix $\mathbf{T}$ describes how the scattered field emitted by the $2^{nd}$ IF (as a "source" IF) is seen by the $1^{st}$ IF (as an "observation" IF). The $1^{st}$ and $3^{rd}$ arguments in the brackets originate from the subscripts of $\mathbf{1}_-^o$ and $\mathbf{2}_+^o$, while the $2^{nd}$ and $4^{th}$ arguments in the brackets are from the superscripts of $\mathbf{1}_-^o$ and $\mathbf{2}_+^o$. In detail, $\mathbf{T}_{12}(o,-;o,+)$ is made of block matrices,

$$\mathbf{T}_{12}(o,-;o,+) = \begin{bmatrix} \{A_{n'm',nm}^{31}\} & \{B_{n'm',nm}^{31}\} & 0 \\ \{B_{n'm',nm}^{31}\} & \{A_{n'm',nm}^{31}\} & 0 \\ 0 & 0 & \{\alpha_{n'm',nm}^{31}\} \end{bmatrix}. \quad (28)$$

If the two IFs are in a *local* medium, the $\{\alpha_{n'm',nm}^{31}\}$ block will vanish as well as the third row and column.

Table 1. The contribution of the scattered field emitted by the $2^{nd}$ IF to the incident field at the $1^{st}$ IF

|  | $2^{nd}$ IF (Inner Region) | $2^{nd}$ IF (Outer Region) |
|---|---|---|
| $1^{st}$ IF (Inner Region) | $\mathbf{T}_{12}(i,-;i,+)$ | $\mathbf{T}_{12}(i,+;o,+)$ |
| $1^{st}$ IF (Outer Region) | $\mathbf{T}_{12}(o,-;o,+)$ | $\mathbf{T}_{12}(o,-;o,+)$ |

We recall that our goal is to study how the scattered field due to the $2^{nd}$ IF can contribute to the incident field to the $1^{st}$ IF. In the above, we have analyzed how the scattered field in the outer region of the $2^{nd}$ IF contributes to the incident field in the outer region of the $1^{st}$ IF. Other combinations (in total three) exist (see Table 1): the coupling between the scattered field in the outer region of the $2^{nd}$ IF (i.e., $\mathbf{2}_+^o$) and the incident field in the inner region of the $1^{st}$ IF (i.e., $\mathbf{1}_+^i$); the coupling between the scattered field in the inner region of the $2^{nd}$ IF (i.e., of $\mathbf{2}_-^i$) and the incident field in the inner (outer) region of the $1^{st}$ IF, i.e., $\mathbf{1}_+^i$ ($\mathbf{1}_-^o$). In terms of matrices, we have,

$$\mathbf{1}_+^i = \mathbf{T}_{12}(i,+;o,+) \cdot \mathbf{2}_+^o, \quad (29)$$
$$\mathbf{1}_-^o = \mathbf{T}_{12}(o,-;i,-) \cdot \mathbf{2}_-^i, \quad (30)$$
$$\mathbf{1}_+^i = \mathbf{T}_{12}(i,+;i,-) \cdot \mathbf{2}_-^i. \quad (31)$$

However, for (29), since the outer region of the $2^{nd}$ IF does not overlap with the inner region of the $1^{st}$ IF, the addition theorem does not apply, the interaction is zero and $\mathbf{T}_{12}(i,+;o,+)$ is a zero matrix. Similar analysis can be done to (30) and (31) as well. Up until this point, we have found the complete relations between the scattered field due to the $2^{nd}$ IF and the incident field upon the $1^{st}$ IF,

$$\begin{pmatrix} \mathbf{1}_+^i \\ \mathbf{1}_-^o \end{pmatrix} = \begin{pmatrix} \mathbf{T}_{12}(i,+;i,-) & \mathbf{T}_{12}(i,+;o,+) \\ \mathbf{T}_{12}(o,-;i,-) & \mathbf{T}_{12}(o,-;o,+) \end{pmatrix} \cdot \begin{pmatrix} \mathbf{2}_-^i \\ \mathbf{2}_+^o \end{pmatrix}$$
$$= \begin{pmatrix} 0 & 0 \\ 0 & \mathbf{T}_{12}(o,-;o,+) \end{pmatrix} \cdot \begin{pmatrix} \mathbf{2}_-^i \\ \mathbf{2}_+^o \end{pmatrix}. \quad (32)$$

This analysis scheme can be routinely applied to other cases in **Fig. A1**(b) and (c) and, finally, reach (13) of the main text.

REFERENCES

[1] J. D. Jackson, *Classical Electrodynamics (Third Edition)*. 1999. Accessed: July 28, 2025. [Online]. Available:

http://archive.org/details/john-david-jackson-classical-electrodynamics-wiley-1999

[2] N. A. Mortensen, "Mesoscopic electrodynamics at metal surfaces: — From quantum-corrected hydrodynamics to microscopic surface-response formalism," *Nanophotonics*, vol. 10, no. 10, pp. 2563–2616, Aug. 2021, doi: 10.1515/nanoph-2021-0156.

[3] Y. Yang et al., "A general theoretical and experimental framework for nanoscale electromagnetism," *Nature*, vol. 576, no. 7786, pp. 248–252, Dec. 2019, doi: 10.1038/s41586-019-1803-1.

[4] A. Xomalis, X. Zheng, A. Demetriadou, A. Martínez, R. Chikkaraddy, and J. J. Baumberg, "Interfering Plasmons in Coupled Nanoresonators to Boost Light Localization and SERS," *Nano Lett.*, vol. 21, no. 6, pp. 2512–2518, Mar. 2021, doi: 10.1021/acs.nanolett.0c04987.

[5] A. Xomalis et al., "Detecting mid-infrared light by molecular frequency upconversion in dual-wavelength nanoantennas," *Science*, vol. 374, no. 6572, pp. 1268–1271, Dec. 2021, doi: 10.1126/science.abk2593.

[6] T. Ding, C. Tserkezis, C. Mystilidis, G. A. E. Vandenbosch, and X. Zheng, "Quantum Mechanics in Plasmonic Nanocavities: from Theory to Applications," *Adv. Phys. Res.*, vol. 4, no. 4, p. 2400144, 2025, doi: 10.1002/apxr.202400144.

[7] C. Zhang et al., "Quantum plasmonics pushes chiral sensing limit to single molecules: a paradigm for chiral biodetections," *Nat. Commun.*, vol. 15, no. 1, p. 2, Jan. 2024, doi: 10.1038/s41467-023-42719-z.

[8] J.-D. Chen et al., "Harnessing plasmon-exciton energy exchange for flexible organic solar cells with efficiency of 19.5%," *Nat. Commun.*, vol. 16, no. 1, p. 3829, Apr. 2025, doi: 10.1038/s41467-025-59286-0.

[9] C. G. Ferreira, C. Sansierra, F. Bernal-Texca, M. Zhang, C. Ros, and J. Martorell, "Bias-Free Solar-to-Hydrogen Conversion in a BiVO4/PM6:Y6 Compact Tandem with Optically Balanced Light Absorption," *ENERGY Environ. Mater.*, vol. 7, no. 4, p. e12679, 2024, doi: 10.1002/eem2.12679.

[10] H. A. Atwater and A. Polman, "Plasmonics for improved photovoltaic devices," *Nat. Mater.*, vol. 9, no. 3, pp. 205–213, Mar. 2010, doi: 10.1038/nmat2629.

[11] A. V. Zasedatelev et al., "A room-temperature organic polariton transistor," *Nat. Photonics*, vol. 13, no. 6, pp. 378–383, June 2019, doi: 10.1038/s41566-019-0392-8.

[12] F. Monticone et al., "Nonlocality in photonic materials and metamaterials: roadmap," *Opt. Mater. Express*, vol. 15, no. 7, pp. 1544–1709, July 2025, doi: 10.1364/OME.559374.

[13] S. Raza, G. Toscano, A.-P. Jauho, M. Wubs, and N. A. Mortensen, "Unusual resonances in nanoplasmonic structures due to nonlocal response," *Phys. Rev. B*, vol. 84, no. 12, p. 121412, Sept. 2011, doi: 10.1103/PhysRevB.84.121412.

[14] N. A. Mortensen, S. Raza, M. Wubs, T. Søndergaard, and S. I. Bozhevolnyi, "A generalized non-local optical response theory for plasmonic nanostructures," *Nat. Commun.*, vol. 5, no. 1, p. 3809, May 2014, doi: 10.1038/ncomms4809.

[15] W. Yan, M. Wubs, and N. Asger Mortensen, "Projected Dipole Model for Quantum Plasmonics," *Phys. Rev. Lett.*, vol. 115, no. 13, p. 137403, Sept. 2015, doi: 10.1103/PhysRevLett.115.137403.

[16] T. Christensen, W. Yan, A.-P. Jauho, M. Soljačić, and N. A. Mortensen, "Quantum Corrections in Nanoplasmonics: Shape, Scale, and Material," *Phys. Rev. Lett.*, vol. 118, no. 15, p. 157402, Apr. 2017, doi: 10.1103/PhysRevLett.118.157402.

[17] C. Mystilidis, G. Vandenbosch, and X. Zheng, "Computational Plasmonics: Boundary Integral Equation Methods in Scattering Problems," in *More Adventures in Contemporary Electromagnetic Theory*, F. Chiadini and V. Fiumara, Eds., Cham: Springer Nature Switzerland, 2025, pp. 195–222. doi: 10.1007/978-3-031-83131-7_9.

[18] C. Mystilidis, " Computational Techniques for Mesoscopic Electromagnetics", PhD dissertation, KU Leuven, 2025 [Online]. Available: https://kuleuven.limo.libis.be/discovery/fulldisplay/lirias4226910/32KUL_KUL:Lirias

[19] M. Fang, Z.-X. Huang, W. E. I. Sha, X. Y. Z. Xiong, and X.-L. Wu, "Full Hydrodynamic Model of Nonlinear Electromagnetic Response In Metallic Metamaterials (Invited Paper)," *Prog. Electromagn. Res.*, vol. 157, pp. 63–78, 2016, doi: 10.2528/PIER16100401.

[20] C. Ciracì et al., "Probing the Ultimate Limits of Plasmonic Enhancement," *Science*, vol. 337, no. 6098, pp. 1072–1074, Aug. 2012, doi: 10.1126/science.1224823.

[21] B. Rivière, *Discontinuous Galerkin Methods for Solving Elliptic and Parabolic Equations*. in Frontiers in Applied Mathematics. Society for Industrial and Applied Mathematics, 2008. doi: 10.1137/1.9780898717440.

[22] K. R. Hiremath, L. Zschiedrich, and F. Schmidt, "Numerical solution of nonlocal hydrodynamic Drude model for arbitrary shaped nano-plasmonic structures using Nédélec finite elements," *J. Comput. Phys.*, vol. 231, no. 17, pp. 5890–5896, July 2012, doi: 10.1016/j.jcp.2012.05.013.

[23] G. Wegner, D.-N. Huynh, N. A. Mortensen, F. Intravaia, and K. Busch, "Halevi's extension of the Euler-Drude model for plasmonic systems," *Phys. Rev. B*, vol. 107, no. 11, p. 115425, Mar. 2023, doi: 10.1103/PhysRevB.107.115425.

[24] L. Li, S. Lanteri, N. A. Mortensen, and M. Wubs, "A hybridizable discontinuous Galerkin method for solving nonlocal optical response models," *Comput. Phys. Commun.*, vol. 219, pp. 99–107, Oct. 2017, doi: 10.1016/j.cpc.2017.05.012.

[25] A. R. Echarri, P. a. D. Gonçalves, C. Tserkezis, F. J. G. de Abajo, N. A. Mortensen, and J. D. Cox, "Optical response of noble metal nanostructures: quantum surface effects in crystallographic facets," *Optica*, vol. 8, no. 5, pp. 710–721, May 2021, doi: 10.1364/OPTICA.412122.

[26] X. Zheng, C. Mystilidis, A. Xomalis, and G. A. E. Vandenbosch, "A Boundary Integral Equation Formalism for Modeling Multiple Scattering of Light from 3D Nanoparticles Incorporating Nonlocal Effects," *Adv. Theory Simul.*, vol. 5, no. 12, p. 2200485, 2022, doi: 10.1002/adts.202200485.

[27] X. Zheng, M. Kupresak, R. Mittra, and G. A. E. Vandenbosch, "A Boundary Integral Equation Scheme for Simulating the Nonlocal Hydrodynamic Response of Metallic Antennas at Deep-Nanometer Scales," *IEEE Trans. Antennas Propag.*, vol. 66, no. 9, pp. 4759–4771, Sept. 2018, doi: 10.1109/TAP.2018.2851290.

[28] Y. Eremin and V. Lopushenko, "Comparative analysis of theories accounting for quantum effects in plasmonic nanoparticles," *J. Quant. Spectrosc. Radiat. Transf.*, vol. 331, p. 109268, Jan. 2025, doi: 10.1016/j.jqsrt.2024.109268.

[29] Y. Eremin, A. Doicu, and T. Wriedt, "Discrete sources method for modeling the nonlocal optical response of a nonspherical particle dimer," *J. Quant. Spectrosc. Radiat. Transf.*, vol. 217, pp. 35–44, Sept. 2018, doi: 10.1016/j.jqsrt.2018.05.026.

[30] G. Miano, G. Rubinacci, A. Tamburrino, and F. Villone, "Linearized Fluid Model for Plasmon Oscillations in Metallic Nanoparticles," *IEEE Trans. Magn.*, vol. 44, no. 6, pp. 822–825, June 2008, doi: 10.1109/TMAG.2007.915835.

[31] X. Zheng, M. Kupresak, V. V. Moshchalkov, R. Mittra, and G. A. E. Vandenbosch, "A Potential-Based Formalism for Modeling Local and Hydrodynamic Nonlocal Responses From Plasmonic Waveguides," *IEEE Trans. Antennas Propag.*, vol. 67, no. 6, pp. 3948–3960, June 2019, doi: 10.1109/TAP.2019.2907807.

[32] C. Mystilidis, X. Zheng, A. Xomalis, and G. A. E. Vandenbosch, "A Potential-Based Boundary Element Implementation for Modeling Multiple Scattering from Local and Nonlocal Plasmonic Nanowires," *Adv. Theory Simul.*, vol. 6, no. 3, p. 2200722, 2023, doi: 10.1002/adts.202200722.

[33] W. Yan, N. A. Mortensen, and M. Wubs, "Green's function surface-integral method for nonlocal response of plasmonic nanowires in arbitrary dielectric environments," *Phys. Rev. B*, vol. 88, no. 15, p. 155414, Oct. 2013, doi: 10.1103/PhysRevB.88.155414.

[34] U. Hohenester and G. Unger, "Nanoscale electromagnetism with the boundary element method," *Phys. Rev. B*, vol. 105, no. 7, p. 075428, Feb. 2022, doi: 10.1103/PhysRevB.105.075428.

[35] L. Huber and U. Hohenester, "A Computational Maxwell Solver for Nonlocal Feibelman Parameters in Plasmonics," *J. Phys. Chem. C*, vol. 129, no. 5, pp. 2590–2598, Feb. 2025, doi: 10.1021/acs.jpcc.4c07387.

[36] X. Zheng, "Dedicated Boundary Element Modeling for Nanoparticle-on-Mirror Structures Incorporating Nonlocal Hydrodynamic Effects," *Adv. Theory Simul.*, vol. 5, no. 12, p. 2200480, 2022, doi: 10.1002/adts.202200480.

[37] R. Ruppin, "Extinction properties of thin metallic nanowires," *Opt. Commun.*, vol. 190, no. 1, pp. 205–209, Apr. 2001, doi: 10.1016/S0030-4018(01)01063-X.

[38] R. Ruppin, "Optical Properties of a Plasma Sphere," *Phys. Rev. Lett.*, vol. 31, no. 24, pp. 1434–1437, Dec. 1973, doi: 10.1103/PhysRevLett.31.1434.

[39] T. Christensen, W. Yan, S. Raza, A.-P. Jauho, N. A. Mortensen, and M. Wubs, "Nonlocal Response of Metallic Nanospheres Probed by Light, Electrons, and Atoms," *ACS Nano*, vol. 8, no. 2, pp. 1745–1758, Feb. 2014, doi: 10.1021/nn406153k.

[40] P. A. D. Gonçalves, T. Christensen, N. Rivera, A.-P. Jauho, N. A. Mortensen, and M. Soljačić, "Plasmon–emitter interactions at the

nanoscale," *Nat. Commun.*, vol. 11, no. 1, p. 366, Jan. 2020, doi: 10.1038/s41467-019-13820-z.

[41] M. H. Eriksen, C. Tserkezis, N. A. Mortensen, and J. D. Cox, "Nonlocal effects in plasmon-emitter interactions," *Nanophotonics*, vol. 13, no. 15, pp. 2741–2751, July 2024, doi: 10.1515/nanoph-2023-0575.

[42] J. Benedicto, R. Pollès, C. Ciracì, E. Centeno, D. R. Smith, and A. Moreau, "Numerical tool to take nonlocal effects into account in metallo-dielectric multilayers," *JOSA A*, vol. 32, no. 8, pp. 1581–1588, Aug. 2015, doi: 10.1364/JOSAA.32.001581.

[43] C. Mystilidis, X. Zheng, and G. A. E. Vandenbosch, "OpenSANS: A Semi-Analytical solver for Nonlocal plasmonicS," *Comput. Phys. Commun.*, vol. 284, p. 108609, Mar. 2023, doi: 10.1016/j.cpc.2022.108609.

[44] T. Dong, Y. Shi, L. Lu, F. Chen, X. Ma, and R. Mittra, "Optical response of cylindrical multilayers in the context of hydrodynamic convection-diffusion model," *J. Appl. Phys.*, vol. 120, no. 12, p. 123102, Sept. 2016, doi: 10.1063/1.4963105.

[45] T. Dong, Y. Shi, H. Liu, F. Chen, X. Ma, and R. Mittra, "Investigation on plasmonic responses in multilayered nanospheres including asymmetry and spatial nonlocal effects," *J. Phys. Appl. Phys.*, vol. 50, no. 49, p. 495302, Nov. 2017, doi: 10.1088/1361-6463/aa9257.

[46] M. Kupresak, X. Zheng, R. Mittra, G. A. E. Vandenbosch, and V. V. Moshchalkov, "Nonlocal response of plasmonic core–shell nanotopologies excited by dipole emitters," *Nanoscale Adv.*, vol. 4, no. 10, pp. 2346–2355, 2022, doi: 10.1039/D1NA00726B.

[47] X. Yan, C. Tserkezis, N. A. Mortensen, G. A. E. Vandenbosch, and X. Zheng, "A Dedicated Modeling Scheme for Nonclassical Optical Response From the Nanosphere-on-Mirror Structure," *IEEE Trans. Microw. Theory Tech.*, vol. 72, no. 4, pp. 2095–2109, Apr. 2024, doi: 10.1109/TMTT.2024.3355983.

[48] N. Kyvelos, N. A. Mortensen, X. Zheng, and C. Tserkezis, "Self-Similar Plasmonic Nanolenses: Mesoscopic Ensemble Averaging and Chiral Light–Matter Interactions," *J. Phys. Chem. C*, vol. 129, no. 7, pp. 3635–3645, Feb. 2025, doi: 10.1021/acs.jpcc.4c07319.

[49] I. J. Bundgaard, C. Nicolaisen. Hansen, P. E. Stamatopoulou, and C. Tserkezis, "Quantum-informed plasmonics for strong coupling: the role of electron spill-out," *JOSA B*, vol. 41, no. 5, pp. 1144–1152, May 2024, doi: 10.1364/JOSAB.512129.

[50] W. C. Chew, *Waves and Fields in Inhomogeneous Media*. IEEE Press, 1995.

[51] P. Halevi, "Hydrodynamic model for the degenerate free-electron gas: Generalization to arbitrary frequencies," *Phys. Rev. B*, vol. 51, no. 12, pp. 7497–7499, Mar. 1995, doi: 10.1103/PhysRevB.51.7497.

[52] S. Raza, M. Wubs, S. I. Bozhevolnyi, and N. A. Mortensen, "Nonlocal study of ultimate plasmon hybridization," *Opt. Lett.*, vol. 40, no. 5, pp. 839–842, Mar. 2015, doi: 10.1364/OL.40.000839.

[53] P. E. Stamatopoulou and C. Tserkezis, "Finite-size and quantum effects in plasmonics: manifestations and theoretical modelling [Invited]," *Opt. Mater. Express*, vol. 12, no. 5, pp. 1869–1893, May 2022, doi: 10.1364/OME.456407.

[54] F. (Frank) Forstmann and R. Gerhardts, *Metal optics near the plasma frequency*. Berlin ; New York : Springer-Verlag, 1986. Accessed: Aug. 08, 2025. [Online]. Available: http://archive.org/details/metalopticsnearp0109fors

[55] F. Forstmann and H. Stenschke, "Electrodynamics at Metal Boundaries with Inclusion of Plasma Waves," *Phys. Rev. Lett.*, vol. 38, no. 23, pp. 1365–1368, June 1977, doi: 10.1103/PhysRevLett.38.1365.

[56] P. J. Feibelman, "Surface electromagnetic fields," *Prog. Surf. Sci.*, vol. 12, no. 4, pp. 287–407, Jan. 1982, doi: 10.1016/0079-6816(82)90001-6.

[57] A. Babaze et al., "Quantum surface effects in the electromagnetic coupling between a quantum emitter and a plasmonic nanoantenna: time-dependent density functional theory vs. semiclassical Feibelman approach," *Opt. Express*, vol. 30, no. 12, pp. 21159–21183, June 2022, doi: 10.1364/OE.456338.

[58] A. D. Rakić, A. B. Djurišić, J. M. Elazar, and M. L. Majewski, "Optical properties of metallic films for vertical-cavity optoelectronic devices," *Appl. Opt.*, vol. 37, no. 22, pp. 5271–5283, Aug. 1998, doi: 10.1364/AO.37.005271.

[59] N. W. Ashcroft and N. D. Mermin, *Solid State Physics*. Cengage Learning, 2011.